\documentclass[12pt]{article}
\usepackage{graphicx}
\usepackage[utf8]{inputenc}  
\usepackage[T1]{fontenc}
\usepackage{amsfonts,amssymb,amsthm,amsmath}
\usepackage[french,english]{babel}
\usepackage{graphics,graphicx}

\usepackage{subfigure}
\usepackage[french,english]{babel}
\usepackage{graphics,graphicx}
\begin{document}
\title{
Analysis of Bell's spaceships "paradox"}
\author{J. M. Vigoureux\\
Institut UTINAM, UMR CNRS 6213,\\Université de
Bourgogne-Franche-Comté, 25030 Besançon Cedex,France.\\
{\em jean-marie.vigoureux@univ-fcomte.fr}\\\\
M. Langlois\\
Besançon, 25000, France\\ 
{\em mj.langlois@wanadoo.fr}}
\maketitle \fontsize{12}{24}\selectfont

\begin{abstract}
We analyse the Bell paradox from the point of view of an inertial referential frame and from the one of an accelerated frame. We calculate the distance between the two accelerated spaceships as it appears in each referential and we show that neither of the two values represent the "real" length of the thread. Calculating this latter gives a clear understanding of the so-called paradox.
\end{abstract}

\text{}\\
\section{Introduction}
\subsection{The "paradox" and its Bell interpretation}
The Bell's spaceships paradox was a thought experiment firstly presented by E. Dewan and M. Beran in 1959 \cite{Dewan} which became more widely known when J. S. Bell gave it a modified version \cite{Bell}. \\
The aim of Bell was to show that the length contraction is a real effect as considered by Fitzgerald, Lorentz and Poincaré. In other words, that there is an \textit{actua}l dynamical nature of Lorentz transformations and  that a moving object consequently do experience a stress.\\
The thought experiment is as follows \cite{Bell}: let us consider two spaceships at rest.
A thin thread, just long enough to span the distance between them,  relies the two spaceships.
Both now start accelerating simultaneously and equally as measured in the frame $\mathcal{R}_O$ of an inertial observer $O$, thus having \textit{the same velocity at all times and consequently remaining the same distance the one from the other in $\mathcal{R}_O$}. \\ 
Following Bell, the two ships being speeding up relative to $O$, the thread is expected to contract more and more. However, this cannot be so since the distance between the two spaceships remains constant at all times. How to explain such a result ? Bell argued
that if the thread does not contract, it is necessarily because "something" prevents it from a contraction.
This "something" is a stress exerted by the electromagnetic ether so that to exactly compensate the contraction. From a microscopic point of view, the electromagnetic intermolecular fields inside a moving object would be distorted causing it to contract, or to become stressed if hindered from doing so, so that it would finally break. \\
In that Bell interpretation, when the velocity becomes too great, the stress can become too important and the thread then breaks. Quoting Bell \cite{Bell}:  \textit{"as the rockets speed up, it
[the thread] will become too short because of its need to Fitzgerald contract and must finally break.
It must break when, at sufficiently high velocity, the artificial prevention of the natural contraction
imposes intolerable stress.}"
(note that this conclusion is not prooved by Bell. According to his own interpretation, it would be in fact possible to imagine that ether effects could modify the internal structure of the object so that to prevent its contraction at any time).\\
This paradox appears to be regarded by some physicists as a proof that a relativistically contracted body experiences a stress. Our aim is to show, as many other authors, that solving the "paradox" does not need such a stress.
In fact, as we shall see, Bell discuss as if the \text{real} thread was in the referential frame of $O$ whereas $O$ only measure its "Lorentz contracted picture" in his referential frame.\\
To analyse the paradox we first come back (in the second part of our introduction) to the difference between  Fitzgerald-Lorenz-Poincaré's and Einstein's interpretations of Lorentz equations. With the Einstein interpretation no stress is involved in the length contraction so that it can play no role in the resolution of the paradox. \\
Our main results agree with most of the previous ones (for example with that of  Petkov \cite{Petkov} and Franklin \cite{Franklin}): we agree with the fact that the thread can break due to unequal accelerations of spaceships (when measured in the referential frame of one of them)
which cause the distance between them to increase. We show that when the thread breaks in the referential frame of spaceships, it also appears to be broken in the inertial frame. We deduce from these results that length contraction plays no role in that situation and that there is no need of invoking some stress exerted by an electromagnetic ether to explain that the thread can break.\\
Some important parts of our calculations are new. We show, for example, that 
the distance between the two spaceships as measured by one of them 
\textit{is not the "proper" length of the thread}. To calculate this latter, we show that it is not possible (but for infinitesimal length) to consider a rest lenght of the thread. We thus explain how to define the length of the thread and we show that whatever may be the elongation of the thread as the spaceships accelerate, its "Lorentz contracted picture" in the referential frame of $O$ remains the same as it was when the spaceships started. This result enlights what happens when the thread breaks.\\
In the conclusion, we discuss some other points of the Bell explanations which are not usually considered.\\

\subsection{Fitzgerald-Lorentz-Poincaré's and Einstein's : two interpretations of length contraction and time dilation}
There has been two different interpretations of relativistic time dilation and length contraction. The first one, that of Fitzgerald, Lorentz and Poincaré, considered that there is an actual physical contraction of a moving object originated from the action of an electromagnetic molecular force exerted by the ether. To this end, Poincaré introduced a sort of pressure (then called the "Poincaré stress") which was thought of as giving its \textit{dynamical} explanation. According to this interpretation, moving objects are \textit{really} contracted by "ether vortices" and their clocks \textit{really} tick at a slower rate. This implies that there is no reciprocity between a moving observer and an observer at rest \cite{Redzic}.
In that interpretation, relativistic effects being due to forces exerted by the ether, an object moving in ether and an object at rest in it are in fact in two quite different situations : one is moving through the ether (and it will consequently be shortened); the other is at rest in the ether and he consequently must keep its size.\\
In Einstein interpretation, Lorentz transformations can be understood as a rotation in the four-dimensional Minkowski space. Exactly as a 3D rotation, a 4D rotation will not induce strain inside an object. For clarity, relativistic effects  can be understood as a \textit{perspective effect} \cite{perpective} as we are used to in current life or in graphic arts. We know perspective due to the distance; Einstein pointed out that there also exists a perspective due to the velocity. Exactly as an object appears to be smaller and smaller as its distance from the observer increases, it also seems smaller (and its time appears to be dilated) as its velocity with respect to the observer increases. The faster the relative velocity, the greater the magnitude of time dilation and of length contraction. We are not used to this latter perspective because in everyday life velocities are very small, exactly as we would be not used to distance-perspective if we could see no further than two or three meters (to complete the parallel, it can be noted that the velocity-perspective has its horizon line which is the velocity of light exactly as the distance-perspective has its horizon line which would represent objects "infinitely" far away).\\
In the perspective we are used to in current life, a distant object is not \textit{really} smaller but it appears to be so because it is \textit{far from the observer}. The effect is not a \textit{property of the object itself, but a property of the relation of the observer to the object } (it is far).\\
In the same way, a moving object is not \textit{really} shortened, but it appears to be so because it is \textit{moving very quickly with respect to the observer}. This effect is not a property of this object itself but a property of its \textit{relation} (\textit{relativity = relation}) to the observer (it is moving with respect to the observer). \\
Contrarily to the Fitzgerald-Lorrentz-Poincaré interpretation, it follows from Einstein's interpretation that time dilation and length contraction are \textit{reciprocal effects}: in fact, when two objects $A$ and $B$  are far from each other, the distance from $A$ to $B$ is the same as that of $B$ to $A$. Their \textit{relation} (distance) to each other is the same. So each one "describes" the other smaller than it really is \cite{Langevin}. In the same way, when two objects $A$ and $B$  are moving relatively to each other, the velocity of $A$ with respect to $B$ is the same as that of $B$ with respect to $A$ and, consequently, relativistic effects being properties of the \textit{relation} (one is moving with respect to the other), each one "sees" a length contraction and a time dilation of the other. \\
Coming back now to our subject, in usual perspective (due to distance) there is no need to consider forces to understand that somebody in the distance appears to be very small. In the same way, in Einstein's interpretation, there is no need to consider forces or stresses to understand that a spaceship will appear shortened in the direction of its motion. It is because there is no need of forces, that Einstein's Special Relativity is not a \textit{dynamics} but a \textit{cinematics}. As there is no need of forces, there is also no need of electromagnetic ether.\\
It is important to emphasize that the Einstein interpretation is a lot simpler than that of Fitzgerald, Lorentz and Poincaré. In fact, this latter interpretation was elaborated at the beginning of the $XX^{th}$ century \textit{well before we know how matter is constituted}. Knowing now that there are many different cohesion forces inside matter, it would have been necessary, within their theory, to prove how and why ether stresses would contract wood, metal, glass, humain body or anything else in the same proportion.
Within Einstein theory, relativistic effects being due to the relation between two objects moving relatively to each other, it is easy to understand that both the spaceship, the wood table inside it or the cosmonaut himself (which are in the same relation to the observer) will appear shortened in the same proportions.\\
It is important to keep in mind these two interpretations because we can still read in scientific popular books false statements as "in his spaceship, the cosmonaut ages more slowly" or "moving clocks in uniform motion ticks at a slower rate than a clock at rest" or "travelling near the speed of light keep you young" (without specify that all these sentences are only valid \textit{from the point of vue of another observer}).
Writing such sentences is to be unaware of the relativity principle which states, with Galileo Galilei, that "movement [rectilinear and uniform] is as nothing". 
So, even if I was travelling at a constant velocity very near that of light (with respect to an inertial referential frame), everything would happen \textit{to me}, in \textit{my} spaceship, \textit{as if I was at rest}, and I would not be able to determine from any experiment whether my spaceship is moving or stationary.\\
To underline all this, let us quote Einstein who, 
on may 1911, wrote a paper in Physikalische Zeitschrift \cite{Einstein} on precisely this problem because he thought that the words time dilation and length contraction could cause confusion. He wrote:
"\textit{the question of whether the Lorentz contraction exists or does not exist in reality is misleading, because it does not exist "in reality" insofar as it does not exist for an observer moving with the object. However, it does exist "in reality" in the sense that it could be detected by physical means by a non-comoving observer"} \cite{Varicak}. The same can be said about usual perspective we are used to in current life.\\
Incidentally, let us also note that keeping in mind these two interpretations can constitute a good epistemological introduction to explain what is a physical theory. Regular discussions are held asking "who discovered the relativity". In France, some physicists and mathematicians still think that it is Poincaré because it was the first to give the correct expression for the Lorentz transforms. In the light of the above discussion, we can clearly see that a physical theory is not a compilation of equations but also involves how to use these equations and how to understand their physical meaning. Poincaré was certainly the first to give the correct Lorentz transformations but his theory was not the Special theory of Relativity. In Einstein and in Lorentz-Poincaré theories, Lorentz equations are the same, but their physical significance and the theory behind them are quite different.

\section{The principle of calculations}
Let us consider two spaceships $M$ and $N$ the velocities and the accelerations of which in a stationary inertial frame $\mathcal{R_O}$ are the same at all times. With such conditions the distance $d$ of each other remains constant in $\mathcal{R_O}$. However, the fact that a spaceship has the same acceleration and the same velocity as another one at all times \textit{in $\mathcal{R_O}$} does not imply that it has the same velocity and the same acceleration in its own instantaneous referential frame (by "instantaneous rest frame" we mean a referential frame which instantaneously accompanies the spaceship that is in which the spaceship is momentarily at rest).\\
- precisely, we will show that the two ends of a thread just long enough to span the distance between the two spaceships
have neither the same velocity nor the same acceleration in the referential frame $\mathcal{R_M}$ of $M$ (or equivalently in that $\mathcal{R_N}$ of $N$).\\
- consequently, the distance between the two spaceships in $\mathcal{R_M}$ must grow as the speed of the spaceships increases causing the thread joining the ships eventually to break.\\
- when the thread breaks, its "Lorentz perspective image" in $\mathcal{R_O}$ of course shows its crack.

\section{The mathematical formalism}
In a previous paper \cite{Langlois14} \cite{Langlois17}, we have defined the notion of  \textit{tangent boost along a worldline}. We also have introduced its matrix $\Lambda$ in the Lie algebra of the orthochronous Lorentz group.  As shown in these papers, these two matrices, which are 
both defined \textit{in the instantaneous rest frame $\mathcal{R}_M$ of $M$}, 
contain the whole dynamical description of any inertial or non inertial system $M$:\\
- the tangent boost $L$ contains the 4-velocity $\textbf{V}$ of $M$ and a basis of its instantaneous physical space $(M,\,E)$.
More precisely,
 the columns of $L$ define a referential frame $\mathcal{R_\mathcal{M}} = (E_0 = \textbf{V},\, E_1,\,E_2,\,E_3)$ of $M$, its first column $\textbf{V}$ being the 4-velocity of $M$
and the three others $(E_1,\,E_2,\,E_3)$ defining an orthonormal basis of its instantaneous \textit{physical space}.\\
- the matrix $\Lambda$ contains the 4-acceleration of $M$ (its first column) and its possible instantaneous Thomas-Wigner rotation \cite{Langlois14} \cite{Langlois17}. \\
As shown in \cite{Langlois14}, the tangent boost  $L$ relating $\mathcal{R_O}$ to $\mathcal{R_M}$ can be directly written down from the knowledge of the 3-velocity $V$ (the spatial velocity) of $M$ only. Noting $I_3$ the unit matrix of size $3$, we in fact have: 
\begin{equation*}\label{boost}
L =L(\gamma, V) =\left(\small
\begin{array}{cc}
 \gamma & \gamma\,{^T}\!V   \\
\gamma\, V &\,\,\, I_3+ \dfrac{\gamma^2}{1+\gamma}V\,{^T}\!V\\
 \end{array}
\right)
\quad \text{with} \quad
V=\left(\small
\begin{array}{cc}
p =\dfrac{dx}{dt} \\
q=\dfrac{dy}{dt}  \\
r=\dfrac{dz}{dt} 
 \end{array}
\right)
\quad \text{and} \quad
\gamma = \frac{dt}{d\tau}
\end{equation*}
where $t$ and $\tau$ are the proper time of $O$ and $M$ respectively.
 The $\Lambda$ matrix can then be deduced from the knowledge of $L$.\\
\begin{equation*} \label{Lambda}
\Lambda = L^{-1}\,\frac{dL}{d\tau}
\end{equation*}
\\
When the $\Lambda$ matrix is known and when it has constant coefficients, the tangent boost $L$ can also be obtained from $\Lambda$ by using
\begin{equation}\label{boost}
L = e^{\tau \,\Lambda} 
\end{equation}
 In the case of an uniformly accelerated system with a rectilinear motion, it is easier to start with the $\Lambda$ matrix.
 Noting the 4-acceleration of $M$ in its instantaneous rest frame $(0,\,a\,=\,\text{Constant},\,0,\,0)$, the $\Lambda$ matrix can be written (there is no Thomas precession) \cite{Langlois14}\cite{Langlois17}
  \begin{equation*}
\Lambda =  \left(
\begin{scriptsize}
\begin{array}{cccc}
 0 & a & 0 & 0 \\
 a & 0 & 0 & 0 \\
 0 & 0 & 0 & 0 \\
 0 & 0 & 0 & 0
\end{array}
\end{scriptsize}
\right)
\end{equation*}
Using eq.(\ref{boost}) we thus obtain
\begin{equation}\label{boost2}
L(\tau )= e^{\tau \,\Lambda}= \left(
\begin{scriptsize}
 \begin{array}{cccc}
                     \cosh a \tau  & \sinh a \tau  & 0 & 0 \\
                     \sinh a \tau & \cosh a \tau & 0 & 0 \\
                     0 & 0 & 1 & 0 \\
                     0 & 0 & 0 & 1
                    \end{array}
 \end{scriptsize}                   
                    \right)
 \end{equation}
 
\section{The worldlines $\mathcal{L}_M$ and $\mathcal{L}_N$ of the spaceships}
In order to define the worldlines of spaceships $M$ and $N$, the best is to use the tangent boost $L(\tau )$ (eq.\ref{boost2}).
Integrating the 4-velocity of $M$ (the first column of $L(\tau)$) with respect to its proper time $\tau$ gives the parametric equation of the worldline $\mathcal{L}_M$ of $M$ in the referential frame of $O$. We get:
\begin{equation}\label{OM2}
\tau \to OM(\tau) = ( \frac{1}{a} \sinh a \tau,\; \frac{1}{a} (\cosh a \tau-1)+ x_0,\,0,\,0)
\end{equation}
where the integration constant has been chosen so that the position of $M$ at start in the inertial reference frame $\mathcal{R_\mathcal{O}}$ is 
 $OM_0 = (0, \,x_0,\, 0,\, 0)$. The worldline $\mathcal{L}_N$ of $N$ is then obtained by translating $\mathcal{L}_M$ by the 4-vector $(0,\,d,\,0,\,0)$:
\begin{equation}\label{ON}
\tau \to ON(\tau) =  OM(\tau)+ (0,\,d\,,0,\,0)=( \frac{1}{a} \sinh a \tau,\, \frac{1}{a} (\cosh(a \tau)-1)+ x_0 + d,\,0,\,0)
\end{equation}
Eqs.(\ref{OM2}) and (\ref{ON}) show that the 4-velocities and the 4-accelerations of $M$ and $N$ in $\mathcal{R}_O$, are the same:
\begin{eqnarray*}
\textbf{V}_N = \textbf{V}_M &=& \frac{dOM}{d\tau}  = \Gamma (1, V) = (\cosh a \tau,\, \sinh a \tau,\,0,\,0)\\
\textbf{A}_N = \textbf{A}_M &=& \frac{d^2 OM}{d\tau^2} = (a \sinh a \tau, a \sinh a \tau,\,0,\,0)
\end{eqnarray*}
So, in $\mathcal{R}_O$, the two spaceships have the same velocity at every instant, and hence they remain at the
same distance $MN=ON(\tau)- OM(\tau) = (0,\,d,\,0,\,0)$ apart at all times (see figs(1) and (2)).

\begin{figure}[h]
\centering
\includegraphics[scale=0.5]{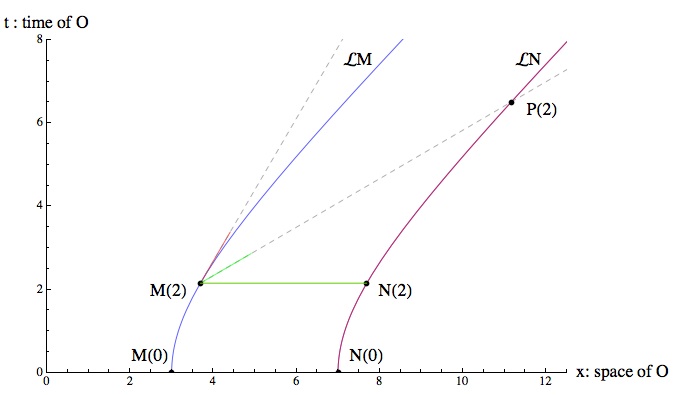}
\caption{This figure shows the worldlines $\mathcal{L_M}$ and $\mathcal{L_N}$ of $M$ and $N$.
The spacetime of $M$ at time $\tau=2$ (that is its time axis and its physical space) is shown in dotted lines.
 $P$ is the intersection of the physical space of $M$ and of the worldline $\mathcal{L_N}$ of $N$. So, $P$ is in the physical space of $M$ and $MP$ consequently is the distance between the spaceships as measured in $\mathcal{R_M}$. This figure has been drawn with $x_0=3, d=4$.}
\end{figure}
\begin{figure}[h]
\centering
\includegraphics[scale=0.5]{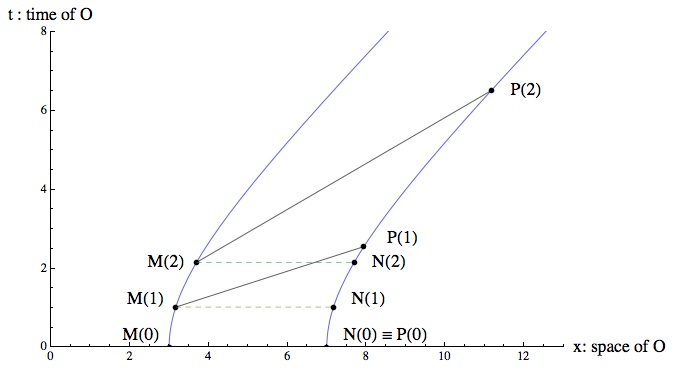}
\caption{This figure shows the evolution of the distance $MP$ between the two spaceships as observed by $M$. $M(\tau) N(\tau)$ (dotted lines) is the distance between the two spaceships at the corresponding times $\tau$ in $\mathcal{R_\mathcal{O}}$ ($O$ judge that the two spaceships remain at the same distance apart). Points $M(\tau)$ and $P(\tau)$ are simultaneous events for $M$. We see that the distance $MP$ between the two worldlines $\mathcal{L}_M$ and $\mathcal{L}_N$ as measured in the referential frame of $M$ increases with time. The point is that $O$ judges that $M$ and $N$ remain at the same distance $d$ ($M(\tau_0)N(\tau_0)=M(\tau_1)N(\tau_1)=M(\tau_2)N(\tau_2)...)$ whereas $M$ judges that it is not the case ($M(\tau_0)P(\tau_0)\,<\,M(\tau_1)P(\tau_1)\,<\,M(\tau_2)P(\tau_2)\,<\,...)$.
The figure as been drawn with $a = 1/3$ and $d = 4$.}
\end{figure}

\section{The distance $MP$ between the two worldlines  $\mathcal{L}_M$ and  $\mathcal{L}_N$ as measured in $\mathcal{R}_M$}
Our aim now is to calculate the distance between the two spaceships in $\mathcal{R}_M$.\\
Let us note $P$ the event of $\mathcal{L}_N$ which is simultaneous with $M$ in $\mathcal{R}_M$ at time $\tau$ (see fig. 1). 
This event is in the physical space of $M$ so that $MP$ represents the distance between the two worldlines (or between the two spaceships) as measured by $M$.
Noting $S$ and $T$  the distance $MP$ in $\mathcal{R}_M$ and the proper time of $P$ on $\mathcal{L}_N$ respectively, $S$ and $T$ can be obtained by writing that $P$ is at the intersection of $\mathcal{L}_N$ and of the physical space $(E_1,\, E_2,\,E_3)$ of $M$, that is to say:
\begin{eqnarray}\label{TS}
\text{$P$ is on the worldline of $N$ : }  \qquad   T &\to& ON(T) \qquad  \text{in $\mathcal{R}_O$}\\
\text{$P$ is in the physical space of $M$ : } \qquad  S &\to& OM(\tau) + S \,E_1(\tau)  \qquad  \text{in $\mathcal{R}_O$}  \nonumber
\end{eqnarray}
The two unknown quantities $T$ and $S$ can thus be obtained by solving the equation
\begin{equation} \label{eq}
OM(\tau) + S\,E_1(\tau)= ON(T)
\end{equation}
Using eqs(\ref{OM2}) and (\ref{ON}) and recalling that $E_1$ is the second column of the tangent boost (\ref{boost2}) so that $E_1=(\sinh a\tau,\, \cosh a \tau,\,0,\,0) $, equation (\ref{eq}) gives
 \begin{eqnarray}\label{tauPP}
(1+a\,S) \, \sinh a \tau &=&\sinh a(T)
  \nonumber\\
 (1+a\,S)\,\cosh a\tau&=&ad+\cosh a(T)
\end{eqnarray}
 Solving then eqs.(\ref{tauPP}), we get $T$ and $S$ in $\mathcal{R}_O$:
 \begin{eqnarray}\label{tauP}
T &=&\tau+ \frac{1}{a}\, (\ln (\sqrt{a^2 d^2 \sinh ^2 a \tau+1}+a d 
   \sinh (a \tau))\nonumber\\
 S&=&\frac{1}{a}\,(\sqrt{a^2 d^2 \sinh ^2 a \tau+1}+a d  \cosh (a \tau)-1)
\end{eqnarray}
Noting $\sinh b(\tau)=a\,d\,\sinh a \tau$ these two results can also be written in the more concise form 
\begin{eqnarray}\label{TS2}
T &=&\frac{a  \tau +b(\tau)}{a}\\
S&=& \frac{\sinh b(\tau)\, \coth a \tau+\cosh b(\tau)-1}{a} \nonumber
\end{eqnarray}
Eq.(\ref{TS}) shows that in $\mathcal{R_O}$, $P$ moves on the same worldline as $N$ but
according to a time law different from that of $N$ ($\tau$ in eq(\ref{ON}) is changed into $T$ in eq.(\ref{TS}))\\
The explicit form of $MP$ at time $\tau$ in $\mathcal{R}_O$ is:
\begin{equation}
 MP_{|\mathcal{R}_O} = S E_1= (S\,\sinh a\tau,\,S \, \cosh a \tau,\,0,\,0) 
   \end{equation}
 where $S$ is given in (\ref{tauP}) and (\ref{TS2}). Its value in $\mathcal{R_M}$ is then
$$MP_{|\mathcal{R}_M} = B^{-1}(\tau).MP_{|\mathcal{R}_O}=(0,\frac{\sqrt{a^2 d^2 \sinh ^2(a \tau )+1}+a d \cosh (a
   \tau )-1}{a},0,0)$$
The figure (2) shows that the distance $MP$ between the two worldlines $\mathcal{L_M}$ and $\mathcal{L_N}$ grows with time in the referential of $M$ whereas it remains $M(\tau)N(\tau)=d=constant$ in that of $O$.\\
It will be usefull in what follows to note that the value of $OP$ in  $\mathcal{R_O}$ is (we use the notation $b$ introduced juste before eq.(9))
\begin{eqnarray*}\label{OPRO}
 OP_{|\mathcal{R}_O} &=&OM+S E_1\\
 &=&(\frac{\sinh a \tau  (a d \cosh a \tau+\cosh b( \tau) )}{a},\frac{\cosh a \tau (a
   d \cosh a \tau +\cosh b( \tau))+a x_0-1}{a},0,0) 
\end{eqnarray*}
\begin{figure}[h]
\centering
\includegraphics[scale=0.5]{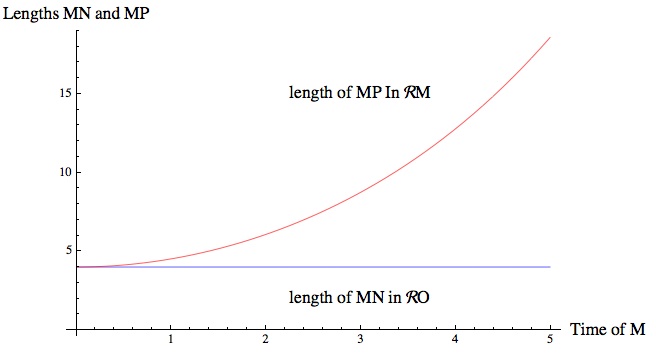}
\caption{Variation of the distance between the two spaceships as measured by $O$ and by $M$ with respect to $\tau$. We see that the distance $MP$ between the two worldlines as measured in the referential frame of $M$ increases continuously with time (so that the thread can break) whereas $MN$ remains constant. This figure has been drawn with $a=\frac{1}{3},x_0=3, d=4$.}
\end{figure}

\section{The velocity and the acceleration of $P$ in $\mathcal{R_M}$}
The 4-velocity $\mathbf{V}_P(\tau)$ of $P$ in $\mathcal{R_O}$ is the same as that of $N$ at time $T(\tau)$, that is $\mathbf{V}_P(\tau)\,=\,\mathbf{V}_N(T(\tau))$. 
It can be written from the first column of $B$ at time $T(\tau)$:
$$\mathbf{V}_P(\tau)=\mathbf{V}_N(T)=\left(\cosh a\,T(\tau),\,\sinh a\,T(\tau),\,0,\,0\right)$$
where $T$ (the proper time of $P$) is given in eqs.(\ref{tauP}) and (\ref{TS2}). 

The 4-acceleration of $P$ in $\mathcal{R_O}$ is consequently
$$\frac{d \mathbf{V}_P}{d \tau}=\frac{d \mathbf{V}_N}{d T}\frac{d T}{d \tau}\quad	\text{with} \quad \frac{d \mathbf{V}_N}{d T}_
{|\mathcal{R}_O} =(0,\,a,\,0,\,0)$$
so that, in $\mathcal{R_M}$
\begin{equation}\label{acc}
\frac{d \mathbf{V}_P}{d \tau}_
{|\mathcal{R}_M} =(0,\,a,\,0,\,0)\, \frac{d T}{d \tau}= \left(0,a\,+\frac{a^2 d  \cosh a \tau}{\sqrt{a^2 d^2 \sinh ^2 a \tau+1}},0,0\right)
\end{equation}
where it can be noted that, in $\mathcal{R}_M $, $\frac{dV_P}{d \tau}$ tends toward $(0,\,2\,a,\,0,\,0)$ when $\tau$ tends toward infinity.

That acceleration is different from the instantaneous acceleration $a$ of the spaceships.
As is well known, this implies, among others, that it is not possible to synchronize clocks  located at different spatial points in a noninertial reference frame.

\section{Will the thread break?}
We see that the velocities and the accelerations of $M$ and $P$ are not the same in $\mathcal{R}_M$ (or in $\mathcal{R}_P$) so that the distance $MP$ between the two spaceships increases with time in $\mathcal{R}_M$ from its initial value $M_0P_0=d$ (figs 2 and 3). This result may be used to explain that the thread must break. 
Sometimes, the reasoning is made using the velocity, sometimes it is made using acceleration saying then truly (following the equivalence principle) that an accelerated reference frame is
locally identical to a reference frame at rest in a gravitational field so that all happens as if an inertial force was applied to the thread.\\
However, although $MP$ do increase with time, this reasoning is not complete. In fact, $MP$ represents the distance between the two worldlines $\mathcal{L}_M$ and  $\mathcal{L}_N$ of $M$ and $P$ (that is the distance between the two spaceships) \textit{as "seen" by $M$} but it does not represent the \text{real} length of the thread itself.\\
To understand this important point, let us ask ourselves the question: what is the lenght of the thread at a given time $\tau$? The velocity $V_P$ of $P$ being different from that $V_M$ of $M$ and consequently the Lorentz factor $\Gamma(V_M)$ being not the same as $\Gamma(V_P)$ what is the real length of $MP$ in $\mathcal{R_O}$?  Is the Lorentz contracted length of $MP$ equal to 
$\dfrac{MP}{\Gamma(V_P)}$ or to $\dfrac{MP}{\Gamma(V_M)} >\dfrac{MP}{\Gamma(V_P)} $ ? To be clear, have a look at fig.(4).
Is  the Lorentz contracted length of $MP$ in $\mathcal{R_O}$ the value
 $MQ_1 < MN$ that we find when using $\Gamma(V_P)$ or 
 $MQ_2 > MN$ that we find with $\Gamma(V_M)$? Because it is a physical attribute of the thread, the length must be clearly and uniquely defined. This is obviously not the case here and this remark is sufficient to conclude that $MP$ cannot be the length of the thread.

\begin{figure}[!h]
\centering
\includegraphics[scale=0.5]{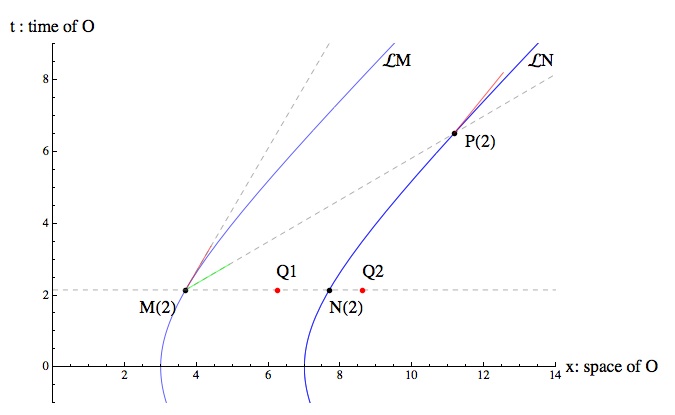}
\caption{This figure shows that the length $MP$ which represents the distance $MP$ between the two worldlines $\mathcal{L}_M$ and $\mathcal{L}_N$ as measured in the referential frame of $M$ does not represent the "proper" lenght of the thread. The velocities being not the same at all points of $MP$ (their values grow continuously form $V_M$ to $V_P$), is the Lorentz contracted of $MP$ in $\mathcal{R}_O$ $MP/\Gamma(V_P) =MQ_1\,<\,MN$ or $MP/\Gamma(V_M) = MQ_2\,>\,MN$? As the previous ones, the figure as been drawn in the case: $a = 1/3,\, x_0=3$ and $d = 4$. } 
\label{}
\end{figure}
Some authors write that the length of the thread must be defined by taking the proper length of the thread that is its length in the referential frame where it is at rest. But obviously such a referential frame does not exist since each point of the thread moves at a different velocity growing from its value $V_M$ at one of its end to $V_P$ at the other end.

The only way to define the proper lenght of the thread, is thus to write:\\
- that each point of the thread must be at rest with respect to its infinitely nearest neighbour, or, in other words, \\
- that each point of the thread must be in the physical space of it infinitely nearest neighbour.\\
So the lenght of the thread can be obtained by constructing the envelope of the family of physical spaces of all points of $MP$. This process is shown on fig.(5)\\
\begin{figure}[!h]
\centering
\includegraphics[scale=0.5]{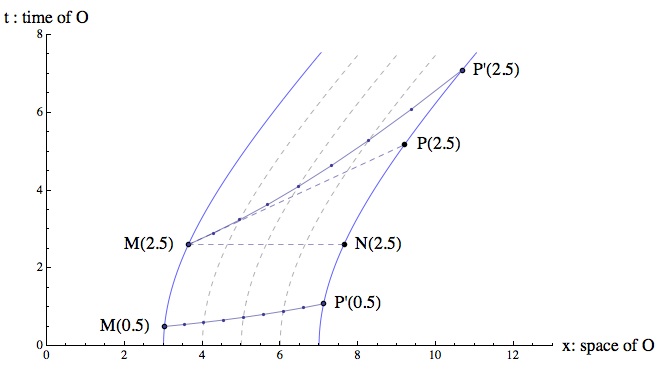}
\caption{This figure shows the thread $MP'$ at two different times. It corresponds to the envelope of the family of physical spaces of all its points between $M$ and $P$. So $MP'$ is the thread istself. $MP$ is its "Lorentz perspective image" in the referential frame of $M$ and $MN$ is its "perspective image" in the referential frame of $O$. As shown in the text, $MP'$ do increase with time but its "image" $MN$ in $\mathcal{R_O}$ remains the same at every time. The figure has been drawn with $n=8$.} 
\label{}
\end{figure}
\indent To construct the envelope of physical spaces of points between $M$ and $P$ and so to represent the thread, we may proceed as follows:\\
- we divide $d = MN$ into $n$ small equal parts $\Delta d$\\  ($MN_1=N_0N_1,\, N_1N_2,\,...,N_i N_{i+1},\,...,N_{n-1},N_n=N_{n-1}N$) and we consider the worldlines $\mathcal{L}_i$ ($i$ varying from $i=0$ at point $M \equiv N_0$ to $i = n$ at point $N$) passing at each $N_i$. \\
- starting then from $N_0 \equiv M$  we calculate the coordinates of $P_1$ and the lenght $M\,P_1$ as
we have calculated $MP$ above. \\
- we then continue from $P_1$ to $P_2$... and so on to the point $P$ so that we get the whole set $P_0P_1,\,P_1P_2,\,...,\,P_{n-1}P_n$.\\
- we then calculate
$$ \sum\limits_{i=1}^{n} P_{i-1}\,P_{i}$$ to get the length of the thread.\\
As shown on figs.(6) and (7) the length $MP'$ of the thread we thus obtain tends toward a well defined value when $n$ tends toward infinity. 
This value is the length of the thread.
$$L=\lim_{n\to\infty}  \sum\limits_{i=1}^{n} P_{i-1}\,P_{i}$$

\begin{figure}[!h]
\centering
\includegraphics[scale=0.4]{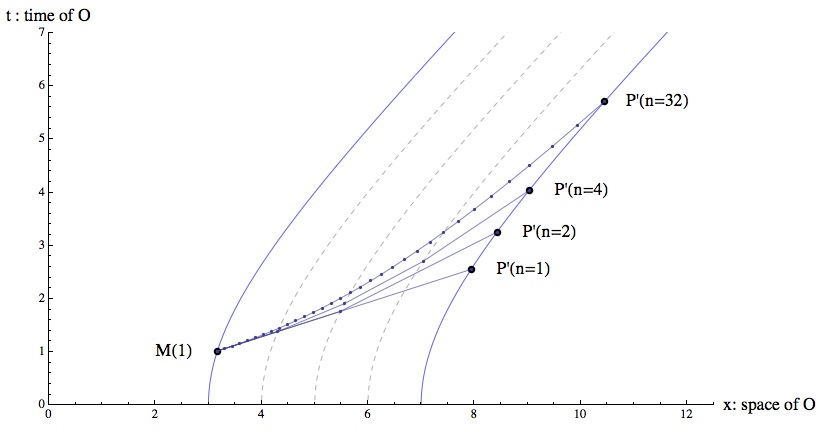}
\caption{To calculate and to represent the length $MP'$ of the thread in $R_0$, we divide the thread in $n$ small elements $\Delta d$. When $n$ tends toward infinity, the length tends toward its real value (see also fig.(7))}
\label{}
\end{figure}

\begin{figure}[!h]
\centering
\includegraphics[scale=0.5]{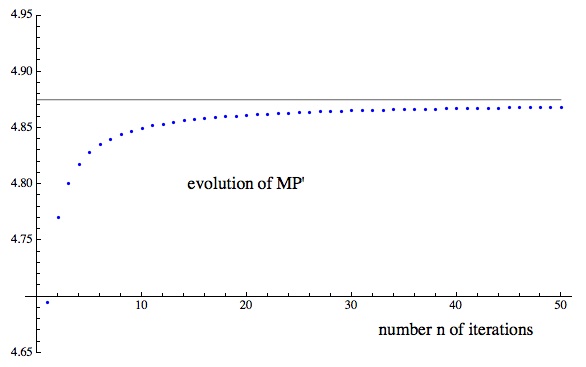}
\caption{The length of the thread as obtained from the above calculations with respect to $n$ at time $\tau=1$ (see fig.6). When $n$ tends toward infinity, the length tends toward its real value (in the present case $4.875$).}
\label{}
\end{figure}

Let us now calculate the length of the thread \textit{as measured in the referential frame of $O$} (that is its Lorentz contracted image). To get it, at each step of the above calculation, we also calculate the velocity $V_i$ of $P_i$ and we deduce the corresponding Lorentz factor $\Gamma(V_i)$. We then calculate
$$  \sum\limits_{i=1}^{n} \frac{P_{i-1}\,P_{i}} {\Gamma(V_i)} \qquad \text{and}  \qquad   \sum\limits_{i=1}^{n} \frac{P_{i-1}\,P_{i}} {\Gamma(V_{i+1})}$$
to get the corresponding point $Q_{1n}$ and $Q_{2n}$ respectively (see fig.4 for the meaning of $Q_1$ and $Q_2$). As expected, $Q_{1n}$ and $Q_{2n}$ tend the one toward the other when $n$ tends toward infinity (see fig.(8)).  This result is the value of the contracted length of the thread as measured in $\mathcal{R_O}$\\
To conclude, all these results show that
with increasing time $\tau$ (and so with increasing velocities of spaceships)\\
- the distance $MP$ between the spaceships and the length of the thread increase in $\mathcal{R_M}$\\
but\\
- the distance between the spaceships remains $MN=d = 4= constant$ in $\mathcal{R_O}$ and \\
- the length of the thread also remains $d=4= constant$ in $\mathcal{R_O}$.\\
So
\textit{whatever may be the distance $MP$ between the two spaceships as measured in $\mathcal{R_M}$, their distance in $\mathcal{R_O}$ remains constant and equal to its initial value $d$}.\\
\begin{figure}[!h]
\centering
\includegraphics[scale=0.5]{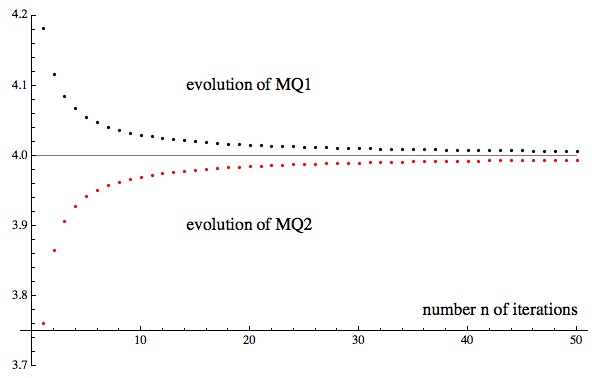}
\caption{To calculate the Lorentz image of the thread in $R_0$, we divide the thread into $n$ small elements $\Delta d$. We then calculate the Lorentz image of each element. Their sum gives points $Q_{1n}$ and $Q_{2n}$. As shown on the figure, $Q_{1n}$ and $Q_{2n}$ tend the one toward the other and toward the value $d=4$ when $n$ tends toward infinity. This result is obtained whatever may be the value of $\tau$ showing that the "image" of the thread in $\mathcal{R_O}$ remains the same ($M_0N_0=d=4$) at all times.}
\label{}
\end{figure}
So, the "real" thread is not $MP$ but $MP'$. $MP$ only represents its "Lorentz contracted picture" in $\mathcal{R_M}$ exactly as $MN$ represents its "picture" in $\mathcal{R_O}$. Neither $MP$ nor $MN$ represent the thread istself. They only are its "perspective pictures" in two different referential frames so that we can foresee that, when the thread will break, its pictures will obviously show its crack (exactly as if somebody is falling at the end of the street, its perspective image will also show that he is falling).

\section{When the thread is broken}
From the viewpoint of an observer located on one of the spaceships, the thread breaks because the spaceships become farther and farther apart as time evolves. Our aim now is to show how this phenomenon will appear to observers $O$ and $M$.
\begin{figure}[!h]
\centering
\includegraphics[scale=0.5]{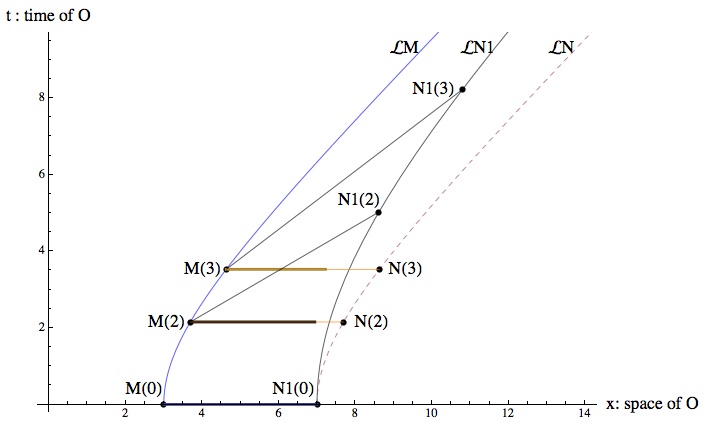}
\caption{ As long as the thread resists to its applied strain, its "image" in $\mathcal{R_O}$ remains the same ($M(\tau)N(\tau)=d=4$) at all times. When it breaks (we suppose here that it breaks at point $N1{0}$ time $t = 0$), it will, from this time, preserve its length so that the worldline of its end is now $\mathcal{L}_{N1}$. So the length of the thread will remain the same in the frame $\mathcal{R_M}$, but its "picture" in $\mathcal{R_O}$ (bold horizontal lines) will appear more and more contracted with increasing time. Consequently, the thread do appear broken, both in $\mathcal{R_M}$ and in $\mathcal{R_O}$.}
\label{}
\end{figure}
When the thread is broken, its end is no more subject to an inertial force due to acceleration and all its parts so have the same velocity as $M$. So its length  $MN_1$ in the referential frame $\mathcal{R_M}$ now remains constant ($MN_1\,=\,(0,\,d,\,0,\,0)$). The worldline $\mathcal{L}_{N1}$ of $N_1$ in $\mathcal{R_O}$ can thus be written (for the sake of clarity we suppose that the thread breaks at its end $N$ at time $\tau=0$)
$$ON_1(\tau)=OM(\tau) + B(\tau).MN_1 = (\frac{(a d+1) \sinh a \tau}{a},\,\frac{(a d+1) \cosh a \tau\,+\,a\,x_0-1}{a},\,0,\,0) $$
 This worldline $\mathcal{L}_{N1}$ of $N_1$ is shown on fig.9 where we can also note that the picture of the thread in $\mathcal{R_O}$ (bold horizontal lines) are smaller than $MN$ so that, as expected, the thread also appears to be broken in $\mathcal{R_O}$. Theses contracted pictures of  $MN_1$ are calculated from
$$\text{contracted length of the thread in }\mathcal{R_O}= \frac{MN_{1\,|\mathcal{R_M}}}{\Gamma (V_M)} $$
These values decreases with time as $V_M$ increases. It can be noted then looking at $M(0)N_1(0)$, $M(2)N_1(2)$ and $M(3)N_1(3)$ on
fig.9 does not show that the length $MN_1$ is constant in the referential frame of $M$. This comes from the fact that fig.9 is drawn in $\mathcal{R}_O$. We must not forget that in this frame, the physical space of $M$ is $(E_1,\, E_2,\,E_3)$ with $E_1=(\sinh a\tau,\, \cosh a \tau,\,0,\,0) $ so that $E_1$ varies with time. To get a clear representation we can represent the same results in the accelerated frame of$M$ (fig.10) which shows that the distance $MN_1$ in  $\mathcal{R_M}$ is clearly constant. That figure is obtained by using
the value of $MP$ in eq.(\ref{acc}).
It can also be obtained by using Rindler coordinates.

\begin{figure}[!h]
\centering
\includegraphics[scale=0.5]{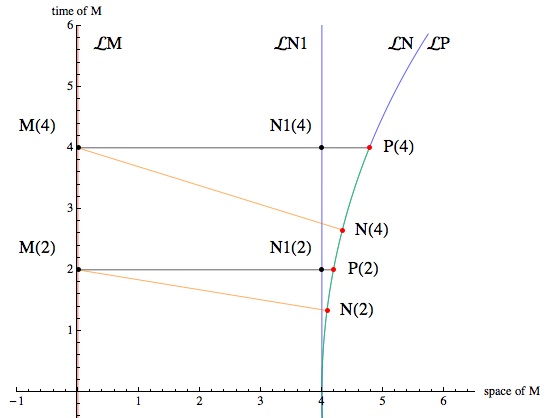}
\caption{The Bell "paradox" can be illustrated in the accelerated referential frame of $M$. $M$ "sees" the other spaceship moving away with increasing time (worldline $\mathcal{L}_N$) whereas the length of the thread (the distance between the two worldlines $\mathcal{L}_M$ and $\mathcal{L}_{N1}$) remains constant in $\mathcal{R}_M$ after its breaking. }
\label{}
\end{figure}

\section{Conclusion } 
One of our main results is to show that the thread is neither the distance $MN$ that is measured in $\mathcal{R_O}$ nor even that $MP$ which is measured in $\mathcal{R_M}$. Because all its points do not have the same velocity before it breaks, the thread has no "proper length" in the common sense. However, its length can be defined by the following conditions:\\
- each point of the thread must be at rest with respect to its infinitely neighbouring points or in other words, \\
- each points of the thread must be in the physical space of its infinitely near neighbour.

So, we have calculated the "proper lengrh" of the thread and we have shown how the thread appears in each referential frame.  We have understood that accelerations of its both ends being not the same, its length constantly increases with time. This, of course explains that the thread can break. However the most interesting result is that, when the thread length increases (that is until it breaks), its "perspective picture" in the inertial frame $\mathcal{R_O}$ remains the same.

From one point of view, there is no paradox here  because we must not confuse the "real" thread, and its "Lorentz pictures" in different referentials: if the thread is broken, its "picture" in any other referential frame shows it is broken; if it is not, its "picture" in any referential shows it is not. The Lorentz transformations only allow to \textit{translate} what happens in a referential frame into the "language" of another one. They imply no change in the objects.
Just as a perspective effect does not induce any real effect on an object, the Lorentz contraction does not induce any strain in the thread.

From another viewpoint, there is clearly a paradox: in fact why should the inertial observer see the thread suddenly breacking whereas he sees it keeping the same apparent length at all times so that he cannot foresee the phenomenon in any way? 
How can he explain it?

Bell suggested resolving the paradox by stating that the thread between the two spaceships
should shorten due to the Fitzgerald-Lorentz-Poincaré contraction but that ether prevents it from doing so by exerting a stress which exactly compensate the contraction. Quoting again his paper \cite{Bell}:  "as the rockets speed up, the thread will become too short because of its need to Fitzgerald contract and must finally break. \textit{It must break when, at sufficiently high velocity, the artificial prevention of the natural contraction imposes intolerable stress.}". We have explained that such an explanation does not wash.
Giving physical properties to ether would in fact lead to extremely difficult problems:\\
- first, because it would break the 
reciprocity between inertial frames in relative uniform motion
and so would raise the question of knowing how to determine our velocity in the ether. Which observer is moving in the ether and which one is not? \\
- second, but not the least: how would it be possible to describe how the ether acts on matter? Answering that question would require weighty assumptions about the constitution of matter.
In fact Fitzgerald, Lorentz and Poincaré elaborated such an interpretation
at the beginning of the $XX^{th}$ century \textit{well before we know 
how matter is constituted}. Knowing now that there are many different cohesion forces inside matter, it would be necessary, within that interpretation, to prove how and why ether stresses would contract any material (wood, metal, glass, humain body or anything else) in the same proportion.
What in fact would happen if we use between the two spaceships different threads made of different materials?
 
So its remains the perspective interpretation which leads to no paradox:
 let us first consider the usual perspective we are used to in current live. Moreover, consider two people holding each end of a rope and moving away progressively from us.
Let us also suppose that we see the length of the rope remaining constant whatever may be its distance from us and that we see it suddenly breaking.
How can we understand this observation since there is apparently no force which could explain it? The reason will be found in noting the lack of perspective:
because of perspective, an object \textit{must} appear smaller and smaller as its distance from the observer increases. So if the length of the thread remains the same although it moves away, it is because the two poeple who hold it are more and more widely apart from each other with distance so applying a force to the thread. The same is true in the case of spaceships: if we see the length of the thread remaining constant 
 whereas it would appear progressively shortened, it is because the two spaceships are more and more widely apart from each other so applying a force to the thread.
  
So, indeed, the thread can break but not for the reason given by Bell:
it breaks because its real lenght increases more and more.\\
At this point, one can ask ourselves to know why Bell's colleagues did not conclude that the thread can break... Quoting Bell : \textit{"This old problem came up for discussion once in the CERN canteen. A distinguished experimental physicist refused to accept that the thread would break, and regarded my assertion, that indeed it would, as a personal misinterpretation of special relativity. We decided to appeal to the CERN Theory Division for arbitration, and made a (not very systematic) canvas of opinion in it. There emerged a clear consensus that the thread would not break!”}. The existence of such a consensus on a bad answer is not without posing its own problems and we can think that the negative answer of CERN Theory Division was not to assert that the thread would not break, but to say that it would not break \textit{because of contracting properties of aether}.

To conclude this study, let us add that the interest of "Bell's paradox" is to show
a manifestation of a \textit{specific} relativistic phenomenon. Not only because it 
does not have a classical analog as underlined by Franklin \cite{Franklin}, but also because it shows an example of a phenomenon which cannot be foresse without noting a lack of perspective.


\begin{thebibliography}{99}
\bibitem{Dewan} E. Dewan et M. Beran, Note on stress effect due to relativistic contraction, 1959, Am. J. Phys. 27, 517-518.
\bibitem{Bell} J. S. Bell: “How to teach special relativity”. 
In Speakable and Unspeakable in Quantum Mechanics 
(Cambridge University Press, Cambridge 1987) pp. 67-68.
\bibitem{Petkov} Vesselin Petkov, \textit{Accelerating spaceships paradox and physical meaning of
length contraction},  arXiv: 0903.5128v1 [physics.class-ph], 30 mars 2009.
\bibitem{Franklin} Jerrold Franklin, \textit{ Lorentz contraction, Bell’s spaceships, and rigid
body motion in special relativity}, arXiv: 0906.1919v3 [physics.class-ph], 8 jan 2010.
\bibitem{Redzic} D. V. Red\v{z}i\'c, Relativistic length agony continued, arXiv: 1005.4623v1 [physics.class-ph], 25 may 2010.
\bibitem{perpective} J. M. Vigoureux, \textit{L'Univers en perspective}, Ellipses (France), 2005.
\bibitem{Langevin} The case of Langevin twins paradox can also be understood in that perpective interpretaton. See J. M. Vigoureux, \textit{L'Univers en perspective}, Ellipses (France), 2005.
\bibitem{Flores} F. Flores, Phys. Educ. 2005n p. 500-503.
\bibitem{Einstein} A. Einstein, "on Ehrenfest paradox. Comment on V. Varicak paper in Physikalische Zeitschrift, vol XII, 1911, p. 509-510.
\bibitem{Varicak}  see also Albert Einstein, Œuvres choisies, 2, Relativité, Seuil-CNRS, p. 80 and Galina Weinstein, a discussion of special relativity, arXiv:1205.0221 [physics.hist-ph], 2012 (or arXiv:1205.0221v1 [physics.hist-ph] 
\bibitem{Langlois14} M. Langlois, J. M. Vigoureux, \textit{Introduction of the Lie group of Lorentz matrix in special relativity. Tangent boost along a worldline and its associated matrix in the Lie algebra. Applications}. ArXiv gr-qc 1411.7254v3, 8 sept. 2016.
\bibitem{Langlois17} M. Langlois, M. Meyer, J. M. Vigoureux, Using tangent boost along a worldline and its associated matrix in the Lie algebra of the Lorentz group, Journal of Modern Physics, 2017, 8, 1190-1212.
\bibitem{Gourgoulhon} E. Gourgoulhon, \textit{Special Relativity in General Frames ; From Particles to Astrophysics}, Springer (Berlin) (2013) or \textit{Relativité restreinte: des particules à l'astrophysique}
EDP Sciences, CNRS Editions (2010).

\end{thebibliography}
\end{document}